 \documentclass[aps,prl,twocolumn,superscriptaddress]{revtex4-1}
  \usepackage{graphicx}
  \usepackage{epstopdf}

\begin{document}

\title{Lie Algebraic Similarity Transformed Hamiltonians for Lattice Model Systems}
\author{Jacob M. Wahlen-Strothman}
\affiliation{Department of Physics and Astronomy, Rice University, Houston, Texas 77005, USA}
\author{Carlos A. Jim\'{e}nez-Hoyos}
\affiliation{Department of Chemistry, Rice University, Houston, Texas 77005, USA}
\author{Thomas M. Henderson}
\author{Gustavo E. Scuseria}
\affiliation{Department of Physics and Astronomy, Rice University, Houston, Texas 77005, USA}
\affiliation{Department of Chemistry, Rice University, Houston, Texas 77005, USA}
\date{ \today}

\begin{abstract}
We present a class of Lie algebraic similarity transformations generated
by exponentials of two-body on-site hermitian operators whose Hausdorff
series can be summed exactly without truncation.
The correlators are defined over the entire lattice and include the
Gutzwiller factor $n_{i\uparrow}n_{i\downarrow}$, and two-site products of density $(n_{i\uparrow} + n_{i\downarrow})$
and spin $(n_{i\uparrow}-n_{i\downarrow})$ operators.
The resulting non-hermitian many-body Hamiltonian can be solved in
a biorthogonal mean-field approach with polynomial computational cost.
The proposed similarity transformation generates locally weighted orbital
transformations of the reference determinant.
Although the energy of the model is unbound,
projective equations in the spirit of coupled cluster theory
lead to well-defined solutions.
The theory is tested on the 1D and 2D repulsive Hubbard model
where we find accurate results across all interaction strengths.
\end{abstract}

\maketitle

\textit{Introduction.}---Hamiltonian similarity transformations are ubiquitous in many areas of
physics, including electronic structure and condensed matter theories, and have been applied in a myriad of contexts~\cite{Boys,
Glazek, Wegner, White, Yanai, Nooijen}. Jastrow-Gutzwiller correlation factors are
also very popular as variational wave functions in quantum Monte Carlo and other
applications~\cite{Fulde,Gutz-review, Baeriswyl2,Sorella,Baeriswyl1,Henderson,Gutz,Neuscamman2,Imada}.
Non-variational solutions have also been
discussed in the literature. Tsuneyuki~\cite{Tsuneyuki} presented a Hilbert
space Jastrow method based on a Gutzwiller factor $\sum_{i}n_{i\uparrow
}n_{i\downarrow }$
and applied it to the 1D Hubbard model,
minimizing its energy variance as in the transcorrelated method~\cite{Handy,
Ten-no,Tsuneyuki2}. Neuscamman \textit{et al.}~\cite{Chan}
proposed
many-body Jastrow correlators, diagonal in the lattice basis, and truncated
them to a subset of sites matching a given pattern; these authors compared
projective solutions with those obtained stochastically via Monte Carlo.

Here, we consider Hamiltonian transformations of the form $e^{-J}He^{J}$ based on hermitian correlators $J
$ built from general two-body products of on-site operators over the entire
lattice.
The transformations here are generated by density (charge), spin, and
Gutzwiller factor correlators, including density-spin crossed terms.
Similar Jastrow-type correlators have been extensively discussed in the
literature but almost always in a variational context~\cite{Gutz-review}. Our
 transformed Hamiltonian is non-hermitian
but can be solved in mean-field via projective equations similar in spirit to those of coupled
cluster theory~\cite{Chan,Bartlett}. In this sense, the model is
an extension that fits under the \textit{generalized} coupled cluster label~\cite{Kutzelnigg,Piecuch,Nooijen2}.
The fundamental difference is that traditional coupled cluster is
formulated with particle-hole excitations out of a reference determinant via
a non-hermitian cluster operator; the present model is constructed with
on-site hermitian correlators.

The main result of this paper is the realization that the Hausdorff series
resulting from the non-unitary similarity transformation
$e^{-J}He^{J}$ can be analytically summed. This result follows from Lie
algebraic arguments~\cite{Gilmore}
 after recognizing that both the Hamiltonian
and the correlator $J$ can be written in the basis of generators of an enveloping
algebra built from on-site operators~\cite{Jinmo}.
Topologically,
our transformation is non-compact and yields a non-hermitian Hamiltonian,
whereas traditional canonical transformations are almost always chosen to be
unitary, thus compact, and preserve hermiticity. There is a mistaken belief
that quantum canonical transformations must be unitary~\cite{Arlen}; this is
not correct even in the linear case~\cite{NUHF}. From this perspective,
traditional coupled cluster exponentiates the shifts of a nilpotent algebra, whose
Hausdorff series truncates at the fourth commutator (for a two-body $H$).
For two-body correlators, our model leads to a renormalized
$N$-body Hamiltonian that produces locally weighted orbital rotations of a
reference state, leading to expectation values between non-orthogonal determinants.
The general theory of enveloping algebras in electronic structure theory, upon
which the present results follow, will be discussed in detail elsewhere \cite{Jinmo}.
Here, we introduce the main mathematical results in a
self-contained manner, touching upon the physical aspects of the model, and
present benchmark applications to the 1D and 2D Hubbard models.

\textit{Theory.}---Consider on-site fermion creation and annihilation operators $c_{i\sigma}^{\dag }$, $c_{i\sigma}$ and on-site number
operators $n_{i\sigma }=c_{i \sigma}^{\dag }c_{i\sigma }$, where $\sigma=\uparrow,\downarrow$.
 The number operators are idempotent ($n_{i\sigma}^{2}=n_{i\sigma}$)
and satisfy
\begin{equation}
\left[ n_{i\sigma },n_{j\sigma'}\right] =0.
\end{equation}%
The elemental fermion operators are eigenoperators (shifts) of the on-site number
operators:
\begin{equation}
\left[ n_{i\sigma},c_{j\sigma'}^{\dag }\right] =\delta _{i\sigma,j\sigma'}c_{j\sigma'}^{\dag }.
\end{equation}%
Let us now define a general two-body correlator,
\begin{equation}
J=\frac{1}{2}\sum_{i\sigma, j\sigma' }\alpha _{i\sigma, j\sigma'}n_{i\sigma}n_{j\sigma'},  \label{J_def}
\end{equation}%
that is chosen hermitian ($J^{\dag }=J$) with real $\alpha $ amplitudes. We
require $\alpha$ to be zero on the diagonal
to exclude one-body operators. The main result of this paper is the
realization that a \emph{global} non-unitary similarity transformation using
the correlator above yields%
\begin{eqnarray}
e^{-J}c_{k\sigma}^{\dag }e^{J} &=& \exp \bigg( -\sum_{j\sigma'}\alpha _{k\sigma,j\sigma'}n_{j\sigma'}\bigg)
c_{k\sigma}^{\dag } \label{Haus} \\
&=& \exp\big(-J_{k\sigma}\big)c^\dagger_{k\sigma}, \nonumber
\end{eqnarray}%
the exponential of a Hermitian one-body operator that commutes with the fermion operator being
transformed. Using this result and its adjoint, we obtain%
\begin{equation}
e^{-J}c_{k\sigma}^{\dag }c_{l\sigma' }e^{J} = e^{-J_{k\sigma}} c_{k\sigma }^{\dag }c_{l\sigma' }e^{J_{l\sigma'}}. \label{onebody}
\end{equation}
Note how the exponentials carry a \emph{local} weight $\alpha$ that depends on the transformed fermion operators. As $J_{k\sigma}$ only consists of on-site number operators, it is a diagonal one-body operator generating a local transformation. The
algebraic derivation is straightforward and can be found in the Supplemental
Material. When acting on a Slater determinant, these one-body
exponentials produce local, non-uniform orbital
rotations.
An orbital rotation, $e^K$, where $K=\sum_{ij}\Lambda_{ij}c^\dagger_ic_j$, acting on a Slater determinant $|\Phi\rangle$ defined by orbital coefficients $C_{ip}$, where $p\in\{1,\dots, N_o\}$ labels the occupied orbitals for $N_o$ occupied states, produces a new unnormalized Slater determinant with transformed coefficients (see eg.~\cite{PHF})
\begin{equation}
e^K|\Phi\rangle=|\Phi'\rangle,\quad C'=e^\Lambda C.
\end{equation}
Using this property, expectation values of Eq. (\ref{onebody}) with $|\Phi\rangle$ can be calculated as transition density matrix elements between non-orthogonal states \cite{blaizot} by applying the local transformations to the reference state
\begin{equation}
\langle\Phi|e^{ -J_{p\sigma}} c_{p\sigma }^{\dag }c_{q\sigma' }e^{J_{q\sigma'}}|\Phi\rangle=\det(S)~\rho_{q\sigma',p\sigma}, \label{density}
\end{equation}
where
\begin{eqnarray}
S &=& (e^{-\Lambda^{p\sigma}}C)^\dagger~(e^{\Lambda^{q\sigma'}}C),\\
\quad \rho &=& (e^{\Lambda^{q\sigma'}}C)~S^{-1}(e^{-\Lambda^{p\sigma}}C)^\dagger. \nonumber
\end{eqnarray}
Here, $\Lambda^{p\sigma}$ is a diagonal matrix containing the set of coefficients for the operator $J_{p\sigma}$.
The evaluation of a single element of the transformed density for a general correlator of the form given in Eq.~(\ref{J_def}) therefore has $O(MN_o^2)$ cost, where $M$ is the size of the basis. This cost can be reduced for some cases, as explained below. The extension of Eq. (\ref{onebody}) for a two-body operator is straightforward, resulting in similar
expressions with one local weight per fermion operator. This can be evaluated as a two-site
dependent transformation on the Slater determinant with the same cost for the evaluation of each element.

The correlator (\ref{J_def}) includes all combinations of two-body on-site operators. The quantities of interest here are
$N_{i}N_{j},S_{i}^{z}S_{j}^{z},N_{i}S_{j}^{z},S_{i}^{z}N_{j},D_{i}+D_j$, where%
\begin{eqnarray}
N_{i} &=&n_{i\uparrow }+n_{i\downarrow }, \label{cartan}\\
S_{i}^{z} &=&n_{i\uparrow }-n_{i\downarrow },  \nonumber \\
D_{i} &=&n_{i\uparrow }n_{i\downarrow }.  \nonumber
\end{eqnarray}%
These two-body operators acting on a reference
modify correlation corresponding to density, spin, and double-occupancy providing flexibility
to improve approximate wavefunctions with poor descriptions of these correlation functions. Here we
seek to add these corrections in an efficient manner via similarity transformation.

The nearest-neighbor Hubbard Hamiltonian
\begin{equation}
H=-t\sum_{\left\langle ij\right\rangle \sigma }c_{i\sigma }^{\dag
}c_{j\sigma } +U\sum_{i}n_{i\uparrow
}n_{i\downarrow }  \nonumber
\end{equation}%
contains at most two-site terms; $\langle ij\rangle$ represents nearest-neighbors, $t$ is the energy
for a particle to hop from one site to a neighboring site, and $U$ is the interaction of two particles on
the same site. Clearly, $J$ as defined in Eq. (\ref{J_def})
commutes with the interaction but not with hopping. The proposed similarity
transformation yields the nonhermitian effective Hamiltonian $\overline H=e^{-J}He^J$,
\begin{eqnarray}
\overline{H}&=&-t\sum_{\langle ij\rangle\sigma }e^{-J_{i\sigma}}c^\dagger_{i\sigma}c_{j\sigma}e^{J_{j\sigma}}+U\sum_{i}n_{i\uparrow }n_{i\downarrow }.
\end{eqnarray}%
The correlator parameters must be optimized and a suitable non-hermitian optimization scheme is needed
as the energy,
$E_J=\langle\overline H\rangle$,
is unbound with respect to $\alpha$.

The Hamiltonian can be treated via  left
projection with the component operators of $J$.
Schr\"{o}dinger's equation is projected into a subspace, as in coupled cluster theory, leading to a system of equations
that determine the unknowns $\alpha_{ij}$~\cite{Chan,Bartlett}
\begin{eqnarray}
\langle n_{i\sigma}n_{j\sigma'}(\overline{H}-\langle\overline H\rangle)\rangle&=&0,\quad \forall~i\sigma\neq j\sigma' \label{Proj}.
\end{eqnarray}
This exactly solves Schr\"{o}dinger's equation projected onto the subspace
$\{n_{i\sigma}n_{j\sigma'}|\Phi\rangle, |\Phi\rangle\}$.
A hermitized variance can be constructed and minimized as in transcorrelation~\cite{Tsuneyuki} and is discussed in the Supplemental Material.
Quantities other than the energy can be calculated via a Lagrangian formulation analogous to that used in coupled cluster theory~\cite{Bartlett},
\begin{equation}
L=\langle \overline H\rangle+\sum_{i\sigma\neq j\sigma'}z_{i\sigma,j\sigma'}\langle n_{i\sigma}n_{j\sigma'}(\overline H -\langle\overline H\rangle)\rangle. \label{lagrange}
\end{equation}
Requiring $\partial L/\partial z_{i\sigma,j\sigma'}=0$ results in Eq. (\ref{Proj}) and $\partial L/\partial\alpha_{i\sigma,j\sigma'}=0$ gives the equations for the linear response amplitudes $z$.
The expectation value of an arbitrary operator $\mathcal{O}$ can then be calculated, including the contributions from the response equations, as
\begin{equation}
\langle\mathcal{O}\rangle_J = \langle \overline\mathcal{O}\rangle
+ \sum_{i\sigma\neq j\sigma'}z_{i\sigma,j\sigma'}\langle n_{i\sigma}n_{j\sigma'}(\overline\mathcal{O} -\langle \overline\mathcal{O}\rangle)\rangle \label{response}
\end{equation}
 where $\overline\mathcal{O}=e^{-J}\mathcal{O}e^J$. (See Ref.~\cite{Bartlett} for more details on response equations.)

\begin{figure}[t]
 \includegraphics[width=3.4in]{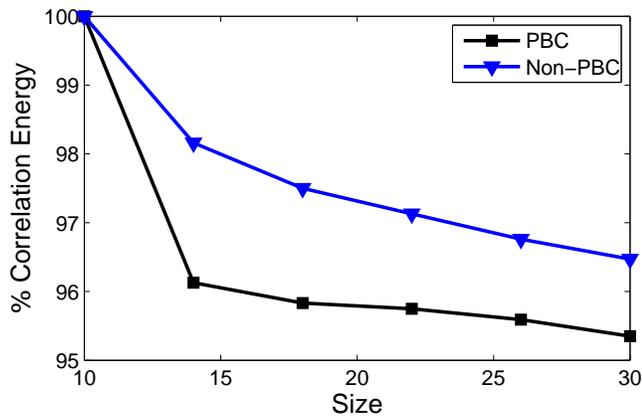}
 \vspace*{-5ex}
\caption{Correlation energy of 8-hole-doped Hubbard chains for $U=2$ with open and closed
boundaries on an RHF reference. DMRG is used to find exact energies for open systems~\cite{ALPS1,ALPS2}.}
\label{fig:dope}
\vspace*{-3ex}
 \end{figure}

\textit{Results.}---We present benchmark calculations for one and two-dimensional Hubbard systems with a
Hartree-Fock Slater determinant reference. Unless
otherwise stated, the calculations include Gutzwiller, density-density, and spin-spin terms, with energy in
units of $t$. The correlation energy is measured with respect to restricted Hartree-Fock (RHF) energies.

 \begin{table}[b]
 \vspace{-2ex}
 \begin{ruledtabular}
\begin{tabular}{rrrrrrr}
\multicolumn{1}{c}{Size} & \multicolumn{1}{c}{$N_o$} &  \multicolumn{1}{c}{U} &   \multicolumn{1}{c}{$E_{RHF}$}  &   \multicolumn{1}{c}{$E_J$} & \multicolumn{1}{c}{$E_{MC}$} & \multicolumn{1}{c}{$\%E_{c}$} \\
\hline
$6\times 6$ & 24 & 4 & -1.0546 & -1.1684 & -1.1853 & 87.06 \\
$6\times 6$ & 24 & 8 & -0.6097 & -0.9845 & -1.0393 & 87.25 \\
$8\times 8$ & 28 & 4 & -1.0078 & -1.0659 & -1.0718 & 90.78 \\
$8\times 8$ & 44 & 4 & -1.0542 & -1.1693 & -1.1858 & 87.75 \\
	\end{tabular}
	\end{ruledtabular}
	\vspace{-2ex}
	\caption{Energy per site and portion of the recovered correlation energy ($E_c$) for 2D, periodic lattices with $N_o$ electrons, an RHF reference wavefunction and released-constraint Monte Carlo ($E_{MC}$) \cite{Zhang1,Zhang2} as the best estimate for the exact result.}
	\label{table1}
	\vspace*{-2ex}
\end{table}

In Fig.~\ref{fig:dope} we compare the correlation energy captured for 8-hole doped systems with periodic and non-periodic boundaries. The theory is most accurate for systems with few particles and open boundary conditions, but as we increase the size of the system, finite size effects are suppressed and we begin to approach the thermodynamic limit while still recovering more than 95\% of the correlation energy. We produce highly accurate results for doped systems and find some reduction in the quality as we approach the thermodynamic limit but still find significant improvements.

We have applied the method to a select set of two-dimensional Hubbard lattices where high quality reference data are available (Table~\ref{table1}). By screening the incorrect double-occupancy with the Gutzwiller factor and incorporating corrections to the correlations in the RHF wavefunction with the spin and density terms, most of the correlation energy is recovered, dramatically improving the results. This supports the method as a cost-effective way to treat larger systems with high accuracy. Calculations
on much bigger lattices are feasible and they will be reported in due course.

\begin{figure}[t]
  \includegraphics[width=3.3in]{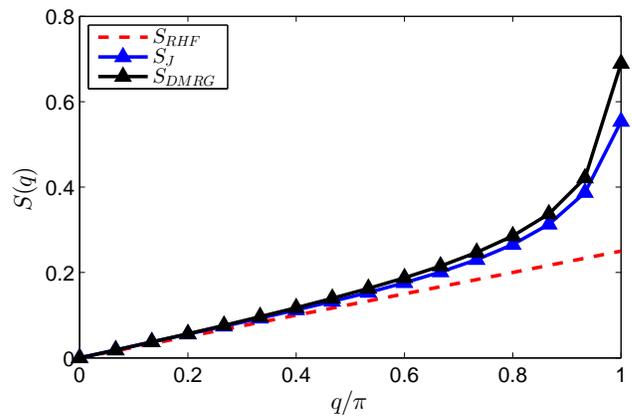}
\vspace*{-2ex}
\caption{The spin-spin correlation function in Fourier space calculated using Eq.~(\ref{response}) ($S_J$) for a 30-site Hubbard ring at half-filling and U=3 compared to the RHF reference and DMRG~\cite{ALPS1,ALPS2}.}
\label{fig:spin}
\vspace*{-3ex}
\end{figure}

We can calculate other significant quantities using Eq.~(\ref{response}) and results agree well with other state-of-the-art methods. Figure~\ref{fig:spin} shows the discrete Fourier transform of the spin-spin correlation function, $S(i)=\langle S^z_0S^z_i\rangle$, for a one-dimensional Hubbard ring. We find that the Jastrow correlator adds most of the correct correlation on an otherwise smooth background. The function is only slightly underestimated at $q=\pi$, unlike the comparatively flat reference, and has the correct long-range decay.

If the wavefunction $|\Phi\rangle$ is a right eigenstate of the transformed Hamiltonian, then $e^J|\Phi\rangle$ is a solution to the original Hamiltonian, and we expect good approximations to $|\Phi\rangle$ and $J$ to have a similar approximate relationship. In order to attest to the power of the method, we compare transformed ($E_J$) and variational energies,
\begin{equation}
E_v=\frac{\langle e^JHe^J\rangle}{\langle e^{2J}\rangle},
\end{equation}
for a 14-site system (Table~\ref{table}). By directly computing overlaps with the exact wavefunction, we can determine how close the correlated wavefunction is to the exact solution. There is strong
agreement in the weak-coupling regime, where the results are of excellent quality, and reasonable
agreement at larger interaction strengths as seen before. This is further supported by the overlap of the reference
and correlated wavefunctions with the true ground state and by the variance per
particle ($0.0038$, 0.0174, and 0.0481 for $U$ of $1$, $2$, and $3$ respectively).
As $e^J|\Phi\rangle$  is close to the true ground state, the Schr\"odinger
equation is nearly satisfied and the energy evaluation using the
transformed Hamiltonian is close to the corresponding variational energy.

 \begin{table}[t]
 \begin{ruledtabular}
\begin{tabular}{rrrrrrr}
\multicolumn{1}{c}{$U$} & \multicolumn{1}{c}{$E_J$} &  \multicolumn{1}{c}{$E_v$} & \multicolumn{1}{c}{$E_{exact}$} & \multicolumn{1}{c}{$E_{RHF}$} & \multicolumn{1}{c}{$|\langle 0|\Phi\rangle|$} & \multicolumn{1}{c}{$|\langle 0|J\rangle|$} \\
\hline
1    &  -14.6983   &    -14.7003   &   -14.7147   &   -14.4758   &   0.9721  &  0.9972 \\
2    &  -11.8486   &    -11.8765   &   -11.9543    &   -10.9758   &   0.8780  &  0.9815\\
3    &  -9.3925     &    -9.5059     &   -9.7488     &   -7.4758     &   0.7100  &  0.9378\\
4    &  -7.4688     &    -7.4745     &   -8.0883     &   -3.9758     &   0.5296  &  0.8711\\
5    &  -5.5017     &    -6.0807     &   -6.8531     &   -0.4758     &   0.3967  &  0.8544\\
6    &  -3.7983     &    -4.8766     &   -5.9165     &    3.0242     &   0.3086  &  0.8437\\
	\end{tabular}
	\end{ruledtabular}
	\vspace{-2ex}
	\caption{Energies and overlaps for the exact $|0\rangle$, RHF $|\Phi\rangle$, and correlated
	wavefunctions $|J\rangle$ for a 14-site ring, where
	$|J\rangle=e^J|\Phi\rangle/|\langle\Phi |e^{2J}|\Phi\rangle|^{\frac{1}{2}}$. }
	\label{table}
	\vspace*{-3ex}
\end{table}

 \begin{table}[b]
 \vspace{-2ex}
 \begin{ruledtabular}
\begin{tabular}{rrrrrrr}
\multicolumn{1}{c}{$N_o$} & \multicolumn{1}{c}{$U$} & \multicolumn{1}{c}{$E_{RHF}$} & \multicolumn{1}{c}{$E_{UHF}$} & \multicolumn{1}{c}{$E_{RJ}$} & \multicolumn{1}{c}{$E_{UJ}$}  & \multicolumn{1}{c}{$E_{ED}$}  \\\hline
14 & 2 & -1.1172 & -1.1644 & -1.1634 & -1.1920 & -1.1982  \\
14 & 4 & -0.7344 & -0.8808 & -0.9018 & -0.9595 & -0.9840 \\
14 & 8 & 0.0313 & -0.5921 & -0.5354 & -0.6691 & -0.7418 \\
16 & 2 & -1.0000 & -1.0973 & -1.0509 & -1.1188 & -1.1261  \\
16 & 4 & -0.5000 & -0.7854 & -0.6931 & -0.8270 & -0.8514 \\
16 & 8 & 0.5000 & -0.4619 & -0.2235 & -0.4873 & -0.5293 \\
\end{tabular}
\end{ruledtabular}
\vspace{-2ex}
\caption{Energies for $4\times 4$ Hubbard lattices with RHF ($E_{RJ}$) and UHF ($E_{UJ}$) references including spin-density correlators ($S^z_iN_j$), compared to exact energies ($E_{ED}$)~\cite{Zhang1,Fano}.}
\label{table3}
\vspace{-2ex}
\end{table}

\begin{figure}[t!]
 \includegraphics[width=3.4in]{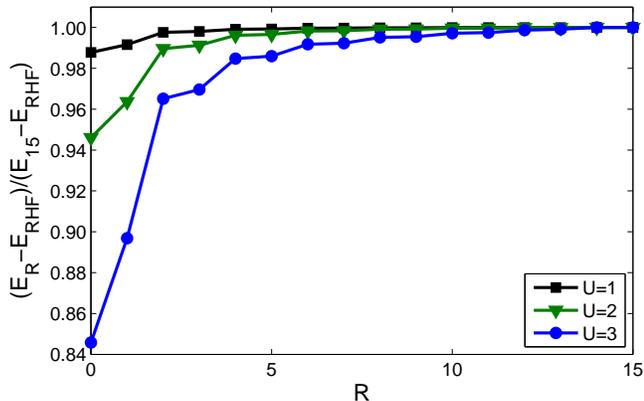}
 \vspace*{-5ex}
\caption{The cumulative fraction of correlation energy captured by limiting the range $R$ of the correlators
compared to the full $R=15$ set for a 30-site ring at half-filling with an RHF reference. $R=0$ includes the Gutzwiller factor.}
\label{fig:range}
 \vspace*{-3ex}
 \end{figure}

For the treatment of larger systems, the cost can be moderated by restricting the correlation
amplitudes to include only local interactions. For sufficiently weak $U$,
the correlations can be limited to short range without significant impact on the quality of the
results. Figure~\ref{fig:range} illustrates this effect. As is clear from the plot, weaker interactions
benefit little from correlation beyond second-nearest neighbors. Truncation at range $R$ results in
$O(MR)$ equations ($O(R)$ for translationally invariant systems) instead of $O(M^2)$, greatly
reducing the computational effort required. Additionally the cost for construction, inversion,
and the determinant evaluation of the overlap matrix in Eq. (\ref{onebody}) can be reduced by a
factor of $M$ via an update of the overlap for each new iteration due to the simple diagonal
structure of the transformations. Truncating the range of the transformation in this manner will restrict the range at which correlation functions calculated with Eq. (\ref{lagrange}) will be accurate, and we believe this approximation is most appropriate in systems where correlations decay rapidly.

There is some reduction in accuracy near half-filling. This can be addressed using a spin-broken reference (Table~\ref{table3}). Whereas the RHF reference has large ionic contributions (zero or double
occupancies), the UHF wavefunction possesses the correct qualitative
antiferromagnetic character near half-filling. As all two-body on-site
correlators are included in our model (Eq.~\ref{cartan}) it is sufficiently
flexible to accommodate the necessary correlations depending on the
nature of the reference. In the case of RHF, the largest contribution to
the correlation energy is typically due to the Gutzwiller factor. Unlike the RHF case, we find a non-zero contribution from spin-density cross terms with a symmetry-broken reference as the up and down orbitals are no longer identical. Results improve significantly with the UHF reference, particularly for large values of U, and we typically recover 80\% or more of the correlation energy.  Additional results are available in the Supplemental Material.

\textit{Conclusions.}---We have presented similarity transformations generated by exponentials
of hermitian on-site operators resulting in a Hausdorff series which can be resummed
and leads to expressions that can be
easily evaluated with polynomial cost. Results from this model
are in very good agreement with the variational
energies, indicating it is a cost effective way of treating wavefunctions of the form $e^J|\Phi\rangle$.
Results for 1D and 2D systems are of high quality with little computational effort. Our method is size extensive, preserves symmetries that commute with $J$, and
is an alternative to variational Monte Carlo sampling with no stochastic error.
The strategy here adopted represents a reasonable approach to optimizing wavefunctions of the form considered in this work without the need to
evaluate the variational energy, which is combinatorial in cost if computed exactly or gains
statistical error if calculated via Monte Carlo.

In extending this idea to non-lattice Hamiltonians, it will be necessary to determine an ``on-site'' basis for the
correlators. In lattice models, we have an obvious choice. The atomic orbital basis may be a good starting point, but better choices might exist.

There are many possible extensions to improve the quality of the results. The theory can incorporate more flexible references such as Hartree-Fock-Bogoliubov or projected BCS wavefunctions with reference optimization.
The marriage of the current on-site correlators and pair coupled cluster doubles (non-hermitian pairing excitation operators in the particle-hole basis)~\cite{pCCD,pqCCD,pCCD2} is promising as they separately address weak and strong
correlation, and is a topic of further study.

\vspace*{1ex}

\textit{Acknowledgments.}---This work was supported by the National Science Foundation
(CHE-1110884). GES is a Welch Foundation Chair (C-0036).

\bibstyle{apsrev4}

\end{document}